\documentclass[conference,compsoc]{IEEEtran}
\ifCLASSOPTIONcompsoc
  \usepackage[nocompress]{cite}
\else
  \usepackage{cite}
\fi
\ifCLASSINFOpdf
\else
\fi
\usepackage{graphicx}
\usepackage{color}
\usepackage{balance}

\IEEEoverridecommandlockouts

\usepackage{tikz}
\newcommand\copyrighttext{%
  \footnotesize 
  R. Teusner, C. Matthies and T. Staubitz, "What Stays in Mind? - Retention Rates in Programming MOOCs," 2018 IEEE Frontiers in Education Conference (FIE), San Jose, CA, USA, 2018, pp. 1-9. doi: \url
  {10.1109/FIE.2018.8658890}. IEEE Xplore: \url{https://ieeexplore.ieee.org/document/8658890}.
  
  \vspace{0.5em}
  
  Copyright \textcopyright 2018 IEEE. Personal use of this material is permitted.
  Permission from IEEE must be obtained for all other uses, in any current or future 
  media, including reprinting/republishing this material for advertising or promotional 
  purposes, creating new collective works, for resale or redistribution to servers or 
  lists, or reuse of any copyrighted component of this work in other works. 
  }
\newcommand\copyrightnotice{%
\begin{tikzpicture}[remember picture,overlay]
\node[anchor=south,yshift=10pt] at (current page.south) {\fbox{\parbox{\dimexpr\textwidth-\fboxsep-\fboxrule\relax}{\copyrighttext}}};
\end{tikzpicture}%
}

\begin{document}
\title{What Stays in Mind? - Retention Rates in Programming MOOCs}

\author{\IEEEauthorblockN{Ralf Teusner, Christoph Matthies, Thomas Staubitz}
\IEEEauthorblockA{Hasso Plattner Institute \\
University of Potsdam, Germany\\
\{firstname.lastname\}@hpi.de}
}

\maketitle

\copyrightnotice

\begin{abstract}
This work presents insights about the long-term effects and retention rates of knowledge acquired within MOOCs. 
In 2015 and 2017, we conducted two introductory MOOCs on object-oriented programming in Java with each over 10,000 registered participants. 
In this paper, we analyze course scores, quiz results and self-stated skill levels of our participants.
The aim of our analysis is to uncover factors influencing the retention of acquired knowledge, such as time passed or knowledge application, in order to improve long-term success.
While we know that some participants learned the programming basics within our course, we lack information on whether this knowledge was applied and fortified after the course's end. 
To fill this knowledge gap, we conducted a survey in 2018 among all participants of our 2015 and 2017 programming MOOCs.
The first part of the survey elicits responses on whether and how MOOC knowledge was applied and gives participants opportunity to voice individual feedback. 
The second part of the survey contains several questions of increasing difficulty and complexity regarding course content in order to learn about the consolidation of the acquired knowledge. 
We distinguish three programming knowledge areas in the survey:
First, understanding of concepts, such as loops and boolean algebra.
Second, syntax knowledge, such as specific keywords. 
Third, practical skills including debugging and coding. 
We further analyzed the long-term effects separately per participant skill group.
While answer rates were low, the collected data shows a decrease of knowledge over time, relatively unaffected by skill level.
Application of the acquired knowledge improves the memory retention rates of MOOC participants across all skill levels.
\end{abstract}

\IEEEpeerreviewmaketitle


\section{Introduction}
Massive Open Online Courses appeal to a broad and diverse audience.
They offer a wide variety of knowledge to anyone, usually free of charge.
These characteristics offer much potential, making it possible to join for anyone interested in a specific topic, regardless of material prosperity, physical access to universities and libraries, or prior knowledge in the respective field.
However, this openness also comes with a downside: a lack of commitment.
Students are not required to invest anything except the click on the enrollment button.
Common ``problems'' of MOOCs are high no-show rates, high dropout rates, and oftentimes missing feedback channels.
These issues are not causing harm to the concept in general, but complicate educators' work of improving learning success and tailoring courses to the audiences' needs.
An approach to tackle these problems is to analyze the data that participants produce on the course platforms and that they share through surveys.
As students signing up and then quitting a course after some time is the most visible and apparent issue, research has focused on predicting, measuring and explaining student dropouts.
What is missing however, is further analysis on the successful case.
What influence do MOOCs have in the long term?
How do students apply their acquired knowledge, and how do they benefit from it?
Research is sparse in this area, due to the previously mentioned lack of commitment limiting the feedback that educators receive. 

To our knowledge, this is also one of the issues that contributes to the challenge of finding a viable and sustainable business model for MOOCs.
Currently, the main motivations for developing and running MOOCs are spreading knowledge, educating the public about specific products, broadcasting desired beliefs and research in general~\cite{renz_opensap_2016, fischer_revenue, belleflamme_economic_2016}.

In order to valuate MOOCs, from an academic as well as a business perspective, it is necessary to prove the lasting effect of the delivered information. 

In this paper, we aim to contribute to the research of knowledge retention within MOOCs, with a focus on programming courses.
We base our conclusions on data collected from two MOOCs and a survey of their participants.

Specifically, we want to answer the following research questions:

\begin{enumerate}
  \item [RQ1.] Does (programming) knowledge fade noticeably over time?
  \item [RQ2.] Does the skill level influence knowledge attrition?
  \item [RQ3.] Does application of the knowledge after the MOOC's end reduce knowledge attrition?
  \item [RQ4.] Does a different time span (1 vs. 3 years) has a measurable effect on knowledge attrition?
\end{enumerate} 

For the following approaches and descriptions, we want to clearly state that whenever we speak of an absolute skill expressed in numbers, we are aware that this numerical value can not reflect the true knowledge, experience and mastery of a topic.

The remainder of this paper is structured as follows:
Section \ref{sec:concept} explains the survey and our rationale behind each question in detail. Section~\ref{sec:related-work} shares related work and relates it to our approach.
Section~\ref{sec:method} shortly outlines our approach, while we present the gathered results in Section~\ref{sec:results}. In the last sections, we conclude our findings, and give an outlook on our upcoming research.


\section{Concept}\label{sec:concept}

Retention of knowledge gained within a MOOC can either be measured with an additional MOOC, in which we could re-identify our prior participants, or with an additional survey.
From the user perspective, it is unlikely that they will enrol in a future iteration of a course they already completed, as it yields little benefit for them.
When offering an advanced course that builds on prior knowledge, it is likely that only participants that completed the first course and have a high interest in the topic will take part, therefore biasing the results.
By offering an additional survey, we hope to reach a broader audience. 
We are aware that also the subset of prior course participants that will answer our survey will be biased, as individuals are naturally more likely to help and answer our questions if they kept us in good memory, and thus probably completed the course or found it in another way appealing.
However, by approaching all prior participants, whether they were no-shows, dropouts, or completed the course with any score, we are confident that we decided for the best possible prerequisite to answer our research questions.

When assessing the effects of a MOOC, we focus on different aspects that could result from course participation or might have occured since course completion:
  
\begin{itemize}
	\item gain of expert knowledge,
	\item refresh and fortification of knowledge,
	\item increase of interest in the problem domain,
	\item correction of misunderstandings,
	\item forgetting of knowledge. 
\end{itemize}

With our survey, we approached a large number of prior participants, therefore we cautiously considered the total amount of questions to ask, the different areas we want to cover and the offered answer options of multiple-choice or multiple-answer questions.
Too many questions bear the risk of scaring participants away, too few questions might leave important areas untouched, leaving gaps when analyzing the responses or reducing precision.
In the end, we decided on 15 questions in total, with one additional free text questions for general feedback and remarks.
In the following, we will further break down our decisions and motivations on the 15 questions that finally made it into the survey.
 
All questions were asked in German, as this was the language used in the course.
 In this paper, we will present the English equivalents of the questions and answers, for the sake of transparency and quality assurance, the original german questions can also be accessed online\footnote{https://rteusner.github.io/fie18/}.
 
The first five questions of the survey are concerned with prior knowledge, self-assessment and perceived valuation of the course.
These questions are intended to give us a general overview over the participants skill level, their prior education, as well as progress in the meantime and valuation of individual course parts.

\begin{enumerate} 
  \item [Q1.] The question ``Did you have programming experience prior to the course?'' separates novices from participants seeking to deepen their knowledge.
This question further allows us to find out, whether participants that initially learned programming with our MOOC lack knowledge that other participants, that adopted their initial understanding from traditional in person classes, have.
The offered answers are ``yes'', ``no'', and ``yes, but it is so long since that I re-learned many things anew''.
  \item [Q2.] The question ``In case you had programming experience prior to the course, where did you learn programming'' aims to uncover differences in different learning ways and stages.
Additionally, albeit in hindsight, it offers us to gain a deeper understanding of our audience.
The offered answers were ``In an online programming course (e.g. Coursera, Udacity, edX, openHPI)'', ``University or school classes'', ``At my workplace'' and ``Self-Study (e.g. books, ...)''.
  \item [Q3.] ``How do you rate your current programming skills'' is typical self-assessment question, with a scale between ``Little to no programming knowledge available (any longer)'', ``basic'', `good'', ``very good'' and ``excellent'' programming skills. This question enables us to to separate students by skill level. It further allows us to notice their self-claimed advance in programming knowledge.
  
  \item [Q4.] ``Did you program since the course ended?'' simply checks whether the participants picked up their (newly) acquired programming skills.
As it is interesting to know, in which context the knowledge was used, we offer the options ``No'', ``Yes, in my spare time (hobby projects)'', ``Yes, as part of my job'', ``Yes, as part of my education (school and university)'', and ``Yes, in self-studies (online courses, books, ...)''.
   \item [Q5.] ``Which parts of the course did you enjoy most in hindsight'' concludes the general questions and aims to give us further insight whether specific focusses of the participant within the course translate to better or worse learning results.
Additionally, it gave us some more feedback about our course parts.
The possible answers were ``the videos'', ``the practical programming exercises'', ``the self-tests'', ``the story'', ``the peer-assessment'', and ``the discussions in the forum''.

\end{enumerate}

After that, we start with multiple-answer questions covering domain knowledge of Java programming.
The questions increase in difficulty and cover different areas.
Additionally, they require different depth of understanding of Java and object-oriented programming.
If not stated otherwise, all answer options were shuffled, so that effects that might skew the results by option placement are prevented.
\begin{enumerate} 
\item [Q6.]	The sixth question asks the participants to ``mark the correct keywords'' from the list of five given options: ``public'', ``common'', ``output'', ``void'', and ``back''.
We purposely mixed the two correct options ``public'' and ``void'' with either valid sounding options ``common'' and ``back'' or otherwise tempting terms like ``output'', which is close to a valid attribute used within the common method call ``\texttt{System.out.println(...);}''.
Both correct options were chosen from the entry point of every Java program, ensuring that also participants that quitted our course after the first video, should be able to at least knowingly pick both of the correct options, even if they fail on sorting out the other options then.
The sixth question thus required solely knowledge without understanding of any concepts or application. 
\item [Q7.]	With the seventh question, we went for one of the key concepts in programming languages, regardless of being them imperative or object-oriented: control structures. 
We started with a question to check whether participants recall basic boolean logic.
To clarify the question, we give them an example where this might be used within their program.
``Which of the given expressions evaluates to true? This could for example be done in an if-statement. Given is the integer variable i with the value 3. ``\texttt{(int i = 3;)}''
There are 5 options given in total in this multiple answer question, with the four correct ones ``\texttt{true}'', ``\texttt{true $\parallel$ false}'', ``\texttt{i = 3}'', and ``\texttt{(i $\leq$ 3 \&\& true) $\parallel$ (i $\leq$ 3 ~\&\& false)} as well as one wrong option ``\texttt{true \&\& false}''.
We implicitly tested participants' understanding of the \texttt{\&\&(AND)}, \texttt{$\|$(OR)} and \texttt{==(equals)} operators and their understanding whether they can solve a more complex boolean equation.
These answers were not shuffled, in order to re-introduce participants with the boolean logic by arranging the statements from simple (just the expression ``true''), to intermediate (the rather simple expression ``\texttt{true $\|$ false}'' and a comparison) to the hard one with AND, a comparison and OR combined.
\item [Q8.] The eighth question tests understanding of a second control structure: the for-loop.
This time, we gave the participants a minimal program, just consisting of a loop with the configuration ``\texttt{(int~i=0; i$\leq$3;i++)}'' which printed out ``I love Java!'' on each iteration.
The question to that was ``How often is I love Java! printed out'', with unshuffled answer options from 0 to 5, and the additional answer ``unlimited times''.
The correct answer, 4 times, requires the participant to understand the program flow, notice that the exit condition for the loop is made up of an ``smaller or equals'' check and requires the understanding how to count with a variable that is initialized with zero.
\item [Q9.] The ninth question returns to assessing knowledge, this time of object-orientation specifics.
The multiple-choice question ``What is the main purpose of an Interface'' requires theoretical knowledge about an advanced concept (at least when regarded from the viewpoint of an introductory course) of OOP.
Participants had to exclusively choose between ``It is a framework to design frontends'', ``It improves performance by better utilization of compiler options'', ``It allows to declare methods to be implemented'', ``It allows to connect to objects: Calls on one of the object are also executed on the other object.'', and ``It allows to restrict the visibility of methods''.
This question has some red herrings that are relatively easy to spot, but at least the visibility restriction answer seems a valid alternative to the correct one, that it declares methods to be implemented.
With the question type set to multiple choice, therefore only allowing only one answer, this is the easiest of the OOP related knowledge questions. 
\item [Q10.] Question ten goes further into asking object-oriented domain knowledge, this time with the more difficult question ``What is polymorphism in object-oriented programming?''.

The possible multiple-answer options are the false one ``Renaming of an object'', the correct one ``The dynamic determination of the actual implementation (dynamic binding) within an inheritance hierarchy offering several potential implementation with identical method signatures'', and the two more false ones ``Calling a static method within an inheritance hierarchy with undefined method signatures'' and ``The instantiation of multiple objects at runtime''.
We tried to give each option a pseudo valid sounding purpose, so that participants have to explicitly decide for each option whether it is true or not.

\item [Q11.] Question eleven combines asking specific Java knowledge and a basic OOP concept: ``How is a inheritance relation created within code?''.
The exclusive multiple-choice options are the correct answer ``By using the keyword extends in the class definition'', as well as the incorrect answers ``By marking the super- and subclass with the keywords child and parent'', ``By using the keyword superclass instead of class'', ``By creating a class Register first, that tells Java which classes inherit from which other ones'' and ``Not at all. Java decides on semantic basis on its own''.

\item [Q12.] Finally, the last OOP knowledge centric question, question twelve, asks ``which statements about abstract classes are true''.
From the four possible answers, the two answers ``Abstract classes can't be instantiated'' and ``Abstract classes are created by adding the keyword abstract in front of the keyword class in the class definition'' are correct.
The two wrong answers, ``Every class is abstract, as it only provides a building instruction for the objects to be created from it'' and ``Abstract classes are created by using the keyword abstractClass instead of the keyword class in the class definition'' , both are close to truth in order to increase difficulty.
Albeit classes primarily  provide building instructions for actual objects, this is not the important difference for abstract classes.
And the second option with regards to keywords is just a slight variation of the correct answer, in order to distinguish participants that maybe recall some concepts, but no longer actively use them, from participants with fortified knowledge.

\end{enumerate}

The remaining three questions aim to test the participants abilities to apply their knowledge. 
We provided code snippets for each question and asked which of the given options has to be inserted at a specified line in order to cause the desired output in question 13 and 14, or what the actual output is for question 15.

\begin{enumerate}
\item [Q13.] Question thirteen shows some code that loops over a given array with a counter variable \texttt{i}.
The loop body adds the value of the array at the current position \texttt{i} to a variable with the identifier \texttt{sum}. 
An if-condition has to be added in order to only sum up those values, that are positive or 0.
The participants have to exclusively choose from six different options.
The correct option is ``\texttt{if(array[i] >= 0)\{}''.
Several other options are offered, for example just checking whether \texttt{i} is greater 0, checking whether i or the array value on position i is smaller than zero, or checking the array element at position \texttt{i+1}, resulting in an array index out of bounds exception.
With this exercise, we intend to check whether participants are able to map the asked requirement to a code construct.
They need to know how they access an array and that it is indexed by zero.
We refrained from giving options with broken syntax, such as trying to access array positions by providing round brackets such as \texttt{array(i)}, because similar, more basic knowledge should be covered already in the knowledge centric questions.
\item [Q14.] Question fourteen asks participants to pick all correct lines of code so that the program prints out all odd numbers between 1 and 10.
The logic to print out the number and to distinguish odd from even numbers is already in place, the participants just have to choose the correct loop condition.
The first correct solution is rather trivial, ``\texttt{for (int i = 0; i < 10; i++)\{}''.
The second correct solution, ``\texttt{for(int i = 1; i < 10; i+=2)\{}'', also prints out the desired numbers and simply skips all even numbers. 
The wrong options just loop over all even numbers or have broken exit conditions such as  ``\texttt{i / 2 == 1}'', ``\texttt{i\%2 < 10}'' or ``\texttt{i\%2==1}''.
Participants need to realize that the correct if-condition is already in place, requiring them to have understood the modulo operator and its usage.
Additionally, they need to notice that also just looping over all odd numbers will create the desired result.
\item [Q15.] Question fifteen presents a complete program and asks participants what the output will be.
The program consists of a loop with five iterations from 0 to 4, printing out the result of a method call each time.
The method being called is named count and receives an integer as an argument.
It simply adds up the passed argument to a sum being kept and returns the current sum value.
The correct output is thus 0, 1, 3, 6, and 10.
This exercise is particularly hard because we have chosen a misleading name for the method count (actually summing values up), and one need to understand loops, method calls, parameters and return values.
Additionally, we used the shorthand form to sum up values, resulting in the line ``\texttt{return result += number;}''.
Other answering options were numbers counting up from 0 to 4, or from 0 to 5, the first numbers of the fibonacci sequence, just some zeros or the option that the program will raise an exception.

\end{enumerate}


\section{Related Work}\label{sec:related-work}
This paper contributes to the research areas Knowledge Assessment and Learning Analytics in MOOCs.
Research in this field seeks to understand the effects of teaching and learning, especially within online learning environments.
While there is plenty of research aiming to predict student's success in ongoing courses, there is relatively little work available on measuring effects, and almost none with regards to long-term effects of MOOCs.

\subsection{Knowledge Retention in Business Trainings}
Outcomes of trainings and effectiveness of learning are often measured in paid business context.
If a training is paid by an employer to train staff for a specific job position, the human resource department usually applies established models (such as Kirkpatrick's 4 levels of evaluation~\cite{kirkpatrick_evaluating_1994}, the ROI methodology by Kirkpatrick and Philipps~\cite{phillips_value_2007}, the Six Sigma approach~\cite{george_lean_2004, smith_six_sigma_1993} as well as others) to measure the outcome.
However, these models do not fit for MOOCs in their current form, as they do not reflect MOOC specifics and are tailored towards their current use case. 
They offer detailed descriptions (reaction of participants to trainings, changes in job workflows, and in the end the return on investment in monetary value in the ROI methodology) or calculations of improvements for production processes (Six Sigma), but they lack measures to e.g. represent initial knowledge. 
Additionally, the models are designed for business situations, mostly staff trainings, which are most often conducted in classroom settings or even smaller scales.

\subsection{Knowledge Retention in Classroom Settings and Theoretical Model}
Retention of knowledge in classroom settings is often explained with the ``learning pyramid'', also called ``cone of experience'' established by Edgar Dale at the National Training Laboratories~\cite{dale_audiovisual_1969}.
While there is dispute about subsequently added factors that should reflect the retention level, it is accepted that more immersion with and application of the learning material improves retention in general.
MOOCs in general offer the basic levels of reading text (via offered articles) up to watch videos (the main course corpus).
Our MOOC in specific also allowed for hands-on-experience via practical programming exercises, therefore further strengthening retention.

Zabrucky and Bays argue that testing improves retention~\cite{zabrucky_testing} in classes.
As MOOCs often offer graded test assignments as well as self tests, this also should strengthen retention within MOOCs.
On the other side, the physical and technically induced distance negatively affects the relationship between teacher and students, as well as the immersion with the content, as all teaching is done in front of a computer.
Existing learning theories should therefore be re-evaluated for their applicability within MOOCs.

\subsection{Retention in Distance Education and MOOCs}
Breslow et al. state that participation as well as performance within MOOCs do not follow the rules by which universities have traditionally organized their teaching. Mainly because MOOCs allow free registration and do not require formal withdrawals. 
They thus ``...do include a large number of students who may not have any interest in completing assignments and assessments.''. 
From their work with data of one of the first huge MOOCs they came to the conclusion, that they ``appreciate what a different animal MOOCs are, and some of the challenges they pose to researchers''~\cite{breslow_studying_2013}.
This suggests that also retention of knowledge might differ from traditional classroom settings.

The MOOC-Maker project~\cite{moocmaker2017}, co-funded by the Erasmus+ program of the European Union, examined retention and attrition rates of MOOCs. 
The consortium of 3 European and 6 Latin American higher education institutes  focused on an analysis of scientific literature that had been published so far and provide a quite complete overview. 
None of the examined literature is analyzing the knowledge retention of the presented content. 
All of the publications dealt with the retention of learners in the course, basically drop-out prediction.
Also Philips' work on retention rates in MOOCs~\cite{phillipsRetention2015}, as well as Bawa's literature overview over retention rates in MOOCs~\cite{bawa_retention_2016} focus on dropout rates when speaking about retention.

Retention rates in the form we viewed them were researched for medical knowledge by Naidr et al.~\cite{naidr_long_term_2004}. 
They found that participants' knowledge retention correlated with their preference of the online course over the classroom course as well as with the number of hours spent with the computer weekly.
Garrison encountered no difference in retention for participants of a distance education program about evidence based medicine compared to a face-to-face training~\cite{garrison_continuing_2007}.
They noticed higher scores for the distance education group for the tests prior to the retention test, however without statistical significance.

\subsection{Survey Participation}
We conducted a survey and expected relatively low response rates, as studies trying to uncover the reasons for dropout report response rates of such surveys between between 12.5\%~\cite{KizilcecAttrition2015} and 1\%~\cite{Whitehill2015Beyond}.
When mailing students after a long period of about one respectively three years, the response rates are likely to strongly drop further.
We therefore kept our expectations low.

\subsection{Self Assessment}
Important parts of our analysis depend on self-stated skill levels. 
As with all metrics reported from a survey, the reliability of self-assessment relies on exact, neutral wording and the offered choices, including the number of options to pick from \cite{jones_optimal_2013, behling_translating_2000}.
Furthermore, tend to avoid the border options and might show some overconfidence \cite{kruger_unskilled_1999}.
This might especially affect our numbers on skill distribution (see Figure~\ref{pic:SkillDistribution}.

\subsection{Additional Sources Reflecting Knowledge}
As the topic of our courses is Object Oriented Programming in Java, we evaluated adding additional information that is only available when solving practical development tasks, such as required time, error rates and issued program runs.  
Previous work has either analyzed this in classroom settings or rather small MOOCs~\cite{yudelson_investigating_2014}.
The data was used to form student models and reflect their knowledge acquisition.
The authors extracted concepts from the developed source code and automatically derive learning paths.
Their use case was thus tracking the current knowledge status, and not documenting long term application of knowledge.
We decided against including a practical exercise, as there was no huge information gain expected, and the required time for students to participate in the survey is likely to double, already with just one exercise.

Apart from the mentioned approaches, it is also possible to incorporate other information to interpret the results, such as demographics like age, the highest degree, or results from previous courses.
According to Morrison and Murphy-Hill, the age and skill of a programmer (proxied by stack overflow reputation in their study) show a positive correlation~\cite{morrison_2013_programming_age}. 
However, as we currently do not have thorough data on these demographics across all participants and we did not  want to ask our participants for such relatively private data just for the sake of assessment without providing direct benefit for them, we omitted these considerations from our study.

Previous work either focused on classroom trainings or regarded retention in terms of dropout prevention, not as actual knowledge retention.
This work presents a specific use case for the important domain of programming and supplies first steps and claims towards generalization within this area in order to start filling the current research gap.

\section{Method}\label{sec:method}
We surveyed the participants of two individual courses from 2015 and 2017 in order to potentially uncover effects caused by an extended timeframe between course participation and current knowledge status.

Invitations to take part in our survey were sent as a course announcement.
The email text encouraged participants to take part in the survey in order to recap their learnings and to support our research.
No benefits or prizes were offered for survey participation.

\subsection{Participants}\label{subsec:participants}

We mailed 20,495 participants of our "Java for Beginners" course in 2015 and 13,609 participants of our "Introduction to Object-Oriented Programming with Java" course in 2017.
The 2015 course resulted in 3,295 participants reaching a graded record of achievement, the 2017 course had 2,124 participants that completed the course with a record of achievement.
The audience covered a wide range of age groups from pupils to retired persons and many others in between with a broad variety of prior knowledge. 
Prior knowledge ranged from senior programmers of structural programming languages such as C or Fortran, to complete beginners with no prior knowledge at all.
The ratio of responses we received was low; lower than we expected. 
We got 66 responses from 2015 ($0.322\%$) and 92 responses from 2017 ($0.676\%$).

\section{Results}\label{sec:results}
The first focus of interest are the results of our survey in general and the success quotes separated by question and participant groups.
Questions Q9-Q12 were asked with identical wording during the runtime of our 2017 course, therefore we also added these mean ``former scores'' of the survey respondents of the 2017 group for better comparability.
The four questions were placed at different places during the 2017 course, questions Q9 and Q10 were asked as a self-test before any content was presented and thus tested prior knowledge, questions Q11 and Q12 were asked in self-tests directly after the respective content unit.   
Outcomes of this general analysis are shown in Table~\ref{table:scores}.

In order to get insights which subgroup of our learners answered to our survey, we compared the distribution of self-stated skill levels.
We further analyzed the self-stated skill levels of participants, that answered the same question in 2017 as well as in the survey of 2018.
69 of the 92 responders from the course of 2017 answered all questions, including the original self-test questions from 2017 which we reintroduced in our survey. 
Of the 69 students who answered all questions, 34 (49\%) stated the same skill level in the survey in 2018 as at the beginning of the course in 2017.
17 participants (25\%) stated a skill reduced by one level, 18 participants (26\%) stated a skill level one option higher.
Stronger variations, either up or down, were not recorded.
All results can be found in Table~\ref{table:skill_levels}.
Noteworthy is, that of the 69 participants who answered all questions, all reached a record of completion (meaning they accessed $> 50\%$ of the course material ), 63 received a graded certificate (reached total scores $> 50\%$).
When adding the 23 students of 2017 who did not answer all questions, we find 4 students that aborted the course, and additional 14 that got a record of completion, but no graded certificate.

\begin{table}[]
\centering
\caption{Mean Survey Scores of Participants from 2015 and 2017}
\label{table:scores}
\begin{tabular}{l|l|l|l}
    & 2017 & 2015 & ``former scores'' from 2017 \\ \hline
Q6  & 0.96 & 0.91 &       \\
Q7  & 0.81 & 0.78 &       \\
Q8  & 0.86 & 0.85 &       \\
Q9  & 0.62 & 0.64 & 0.35  \\
Q10 & 0.66 & 0.67 & 0.32  \\
Q11 & 0.91 & 0.80 & 0.99  \\
Q12 & 0.84 & 0.72 & 0.94  \\
Q13 & 0.84 & 0.80 &       \\
Q14 & 0.66 & 0.63 &       \\
Q15 & 0.66 & 0.50 &      
\end{tabular}
\end{table}

\begin{table}[]
\centering
\caption{Self-Stated Skill-Levels of the 2017 Group in the Current Survey and Changes from their Skill Level Stated During the Course}
\label{table:skill_levels}
\begin{tabular}{r|r|r|r|r}
\multicolumn{1}{l|}{Skill Level} & \multicolumn{1}{l|}{Total: 69} & \multicolumn{1}{l|}{Same: 34} & \multicolumn{1}{l|}{+1: 18} & \multicolumn{1}{l}{-1: 17} \\ \hline
(no knowledge) 0                                & 5 ( 7\%)                       & 1                             & -                           & 4                          \\
(basic knowledge) 1                                & 16 (23\%)                      & 11                            & 3                           & 4                          \\
(good knowledge) 2                                & 36 (52\%)                      & 18                            & 10                          & 8                          \\
(very good knowledge) 3                                & 9 (13\%)                       & 3                             & 5                           & 1                          \\
(excellent knowledge) 4                                & 1 ( 1\%)                       & 1                             & 0                           & -                         
\end{tabular}
\end{table}

From the 18 students that self-stated an improved skill level, 15 (88\%) answered that they programmed in the meantime. 
There are no specific accumulations of answers towards work, hobby, school or self-study visible (all have between 5 and 8 mentions). 
Of the 17 students that answered wit a lower self-stated skill level, 6 answered that they did not program in the meantime (35\%).
All participants that lowered their skill level from ``basic'' to ``not existent (any longer)'' answered they did not program in the meantime.
Of those who programmed in the meantime but lowered their skill level, there is a slight accumulation towards the answer "programmed as a hobby" with 6 mentions opposed to 3 mentions of self-studies, 3 mentions of school and one mention of work.

\begin{figure}
\begin{center}
    \includegraphics[width=1.0\columnwidth]{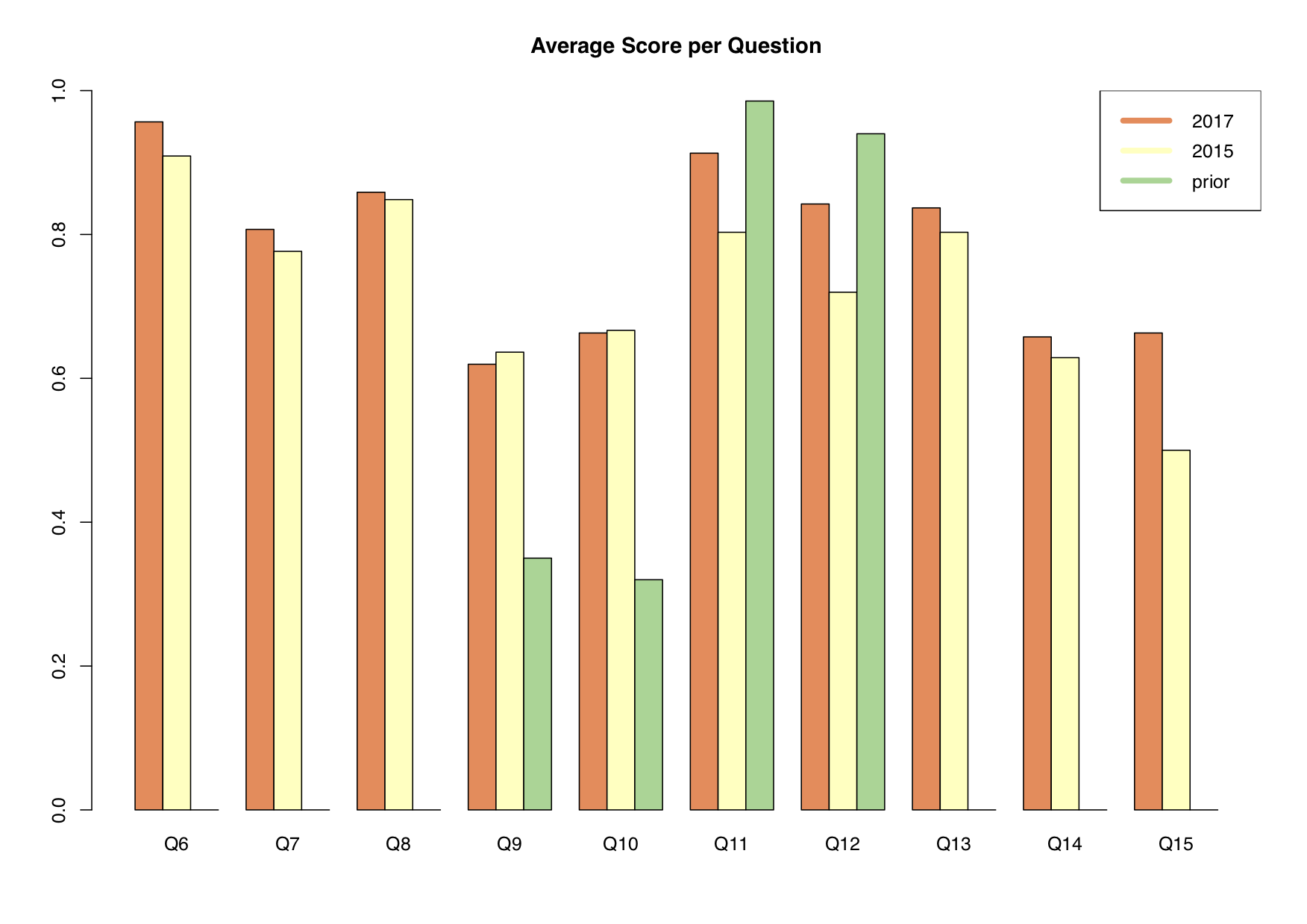}
    \caption{Mean Score per Question and Group}
    \label{pic:AverageScores}
\end{center}
\end{figure}

When having a closer look onto the scores, we get the means vizualized in Figure~\ref{pic:AverageScores} for the graded questions Q6 to Q15.

We further separated the mean scores by skill level, the results can be found in Tables ~\ref{table:averageScoresPerQuestionAndSkillLevel2015} and~\ref{table:averageScoresPerQuestionAndSkillLevel2017}.
We further calculated the deltas for the corresponding mean scores of 2017 and 2015, to approach research question RQ2 (see Table~\ref{table:scoreDeltas}).
Conclusions should be drawn carefully from that table, as the number of students contributing to the respective means is low, especially for skill level 0 (no knowledge) and 4 (expert).
For this reason, also the standard deviations are comparatively high, and not given as the group of skill level 4 in 2017 had just one user.

\begin{table}[]
\centering
\caption{Mean Scores Per Question and Skill Level(SL) 2015}
\label{table:averageScoresPerQuestionAndSkillLevel2015}
\begin{tabular}{@{}l|lllll@{}}
2015        & SL0  {\scriptsize(stdev)}&  SL1 {\scriptsize(stdev)} &  SL2 {\scriptsize(stdev)} &  SL3  {\scriptsize(stdev)} &  SL4  {\scriptsize(stdev)}\\ \hline
Q6  & 0.62 {\scriptsize(0.48)}         & 0.89 {\scriptsize(0.27)}         & 0.92 {\scriptsize(0.19)}         & 0.97   {\scriptsize(0.13)}       & 1.00   {\scriptsize(0.00)}       \\
Q7  & 0.37 {\scriptsize(0.32)}         & 0.64 {\scriptsize(0.35)}         & 0.83 {\scriptsize(0.28)}         & 0.92   {\scriptsize(0.15)}       & 0.92   {\scriptsize(0.14)}       \\
Q8  & 0.25 {\scriptsize(0.50)}         & 0.78 {\scriptsize(0.43)}         & 0.92 {\scriptsize(0.28)}         & 0.94   {\scriptsize(0.25)}       & 1.00   {\scriptsize(0.00)}       \\
Q9  & 0.00 {\scriptsize(0.00)}         & 0.50 {\scriptsize(0.51)}         & 0.76 {\scriptsize(0.44)}         & 0.75   {\scriptsize(0.45)}       & 0.67   {\scriptsize(0.58)}       \\
Q10 & 0.50 {\scriptsize(0.58)}         & 0.67 {\scriptsize(0.49)}         & 0.68 {\scriptsize(0.48)}         & 0.69   {\scriptsize(0.45)}       & 0.67   {\scriptsize(0.58)}       \\
Q11 & 0.00 {\scriptsize(0.00)}         & 0.83 {\scriptsize(0.38)}         & 0.80 {\scriptsize(0.41)}         & 0.94   {\scriptsize(0.25)}       & 1.00   {\scriptsize(0.00)}       \\
Q12 & 0.13 {\scriptsize(0.25)}         & 0.69 {\scriptsize(0.35)}         & 0.78 {\scriptsize(0.41)}         & 0.78   {\scriptsize(0.41)}       & 0.83   {\scriptsize(0.29)}       \\
Q13 & 0.25 {\scriptsize(0.50)}         & 0.67 {\scriptsize(0.49)}         & 0.88 {\scriptsize(0.33)}         & 0.94   {\scriptsize(0.25)}       & 1.00   {\scriptsize(0.00)}       \\
Q14 & 0.00 {\scriptsize(0.00)}         & 0.56 {\scriptsize(0.42)}         & 0.64 {\scriptsize(0.34)}         & 0.88   {\scriptsize(0.22)}       & 0.50   {\scriptsize(0.00)}       \\
Q15 & 0.00 {\scriptsize(0.00)}         & 0.28 {\scriptsize(0.46)}         & 0.72 {\scriptsize(0.46)}         & 0.50   {\scriptsize(0.52)}       & 0.67   {\scriptsize(0.58)}
	\\ 
\hline
Sum &  2.13  & 6.50  & 7.93 & 8.30  & 8.25 \\	
\hline
{\scriptsize\#users} & 4          & 18          & 25          & 16          & 3         
\end{tabular}
\end{table}

\begin{table}[]
\centering
\caption{Mean Scores Per Question and Skill Level(SL) 2017}
\label{table:averageScoresPerQuestionAndSkillLevel2017}
\begin{tabular}{@{}l|lllll@{}}
2017       & SL0 {\scriptsize(stdev)} &  SL1 {\scriptsize(stdev)} &  SL2 {\scriptsize(stdev)} &  SL3 {\scriptsize(stdev)} &  SL4  {\scriptsize(stdev)} \\ \hline
Q6  & 0.90 {\scriptsize(0.22)} & 0.89 {\scriptsize(0.21)} & 1.00 {\scriptsize(0.00)} & 0.94 {\scriptsize(0.17)} & 1.00   \\
Q7  & 0.50 {\scriptsize(0.35)} & 0.81 {\scriptsize(0.33)} & 0.86 {\scriptsize(0.24)} & 0.97 {\scriptsize(0.08)} & 1.00 \\
Q8  & 1.00 {\scriptsize(0.00)} & 0.72 {\scriptsize(0.46)} & 0.92 {\scriptsize(0.28)} & 1.00 {\scriptsize(0.00)} & 0.00 \\
Q9  & 0.40 {\scriptsize(0.55)} & 0.56 {\scriptsize(0.51)} & 0.78 {\scriptsize(0.42)} & 0.89 {\scriptsize(0.33)} & 0.00 \\
Q10 & 0.60 {\scriptsize(0.55)} & 0.61 {\scriptsize(0.50)} & 0.75 {\scriptsize(0.44)} & 0.89 {\scriptsize(0.33)} & 0.00 \\
Q11 & 0.80 {\scriptsize(0.45)} & 0.89 {\scriptsize(0.32)} & 0.97 {\scriptsize(0.17)} & 1.00 {\scriptsize(0.00)} & 1.00 \\
Q12 & 0.60 {\scriptsize(0.55)} & 0.83 {\scriptsize(0.30)} & 0.93 {\scriptsize(0.24)} & 0.94 {\scriptsize(0.17)} & 0.50 \\
Q13 & 0.60 {\scriptsize(0.55)} & 0.83 {\scriptsize(0.38)} & 0.89 {\scriptsize(0.32)} & 0.89 {\scriptsize(0.33)} & 1.00 \\
Q14 & 0.20 {\scriptsize(0.45)} & 0.72 {\scriptsize(0.39)} & 0.71 {\scriptsize(0.37)} & 0.78 {\scriptsize(0.36)} & 0.50 \\
Q15 & 0.20 {\scriptsize(0.45)} & 0.61 {\scriptsize(0.50)} & 0.72 {\scriptsize(0.45)} & 0.89 {\scriptsize(0.33)} & 0.00 \\ 
\hline
Sum &  5.80 & 7.47 & 8.53 & 9.19 & 5.00 \\
\hline
{\scriptsize\#users} & 5          & 18          & 36          & 9          & 1 
\end{tabular}
\end{table}

\begin{table}[]
\centering
\caption{Deltas in Mean Scores Between 2017 and 2015 (Result: MeanScore2017 - MeanScore2015)}
\label{table:scoreDeltas}
\begin{tabular}{l|lllll}
Deltas & SL0 &  SL1 &  SL2 &  SL3 &  SL4 \\ \hline
Q6     & 0,28 & 0,00  & 0,08 & -0,02 & 0,00  \\
Q7     & 0,13 & 0,17  & 0,03 & 0,05  & 0,08  \\
Q8     & 0,75 & -0,06 & 0,00 & 0,06  & -1,00 \\
Q9     & 0,40 & 0,06  & 0,02 & 0,14  & -0,67 \\
Q10    & 0,10 & -0,06 & 0,07 & 0,20  & -0,67 \\
Q11    & 0,80 & 0,06  & 0,17 & 0,06  & 0,00  \\
Q12    & 0,48 & 0,14  & 0,15 & 0,16  & -0,33 \\
Q13    & 0,35 & 0,17  & 0,01 & -0,05 & 0,00  \\
Q14    & 0,20 & 0,17  & 0,07 & -0,10 & 0,00  \\
Q15    & 0,20 & 0,33  & 0,00 & 0,39  & -0,67 \\	
\hline
Sum &  3.67 & 0.97  & 0.60   & 0.90  & -3.25 

\end{tabular}

\end{table}

Correlating the current self-stated skill levels against the unweighted scores of the graded questions, yielded a weak positive correlation of $0.35$ ($p <$ 0.005, test statistic: Pearson's product-moment correlation).
The correlation between the prior self-stated skill levels during the course in 2017 and the unweighted survey scores is also weak and positive, however a little bit higher, namely $0.38$ ($p <$ 0.005).
When correlating the prior course scores with the survey scores, we found no correlation at all ($c = 0.00$).

Calculating the total mean survey scores for participants depending whether they programmed in the meantime or not, yielded a mean of $6.44$ for students that did not program, and a mean of $8.47$ for students that programmed in the meantime, for the 2017 survey group.
For the 2015 survey group, the scores showed a larger gap: a mean score of $3.47$ for non programming students and a mean of $7.32$ for students that programmed in between.  

\subsection{Discussion}\label{subsec:discussion}
\begin{figure}
\begin{center}
    \includegraphics[width=1.0\columnwidth]{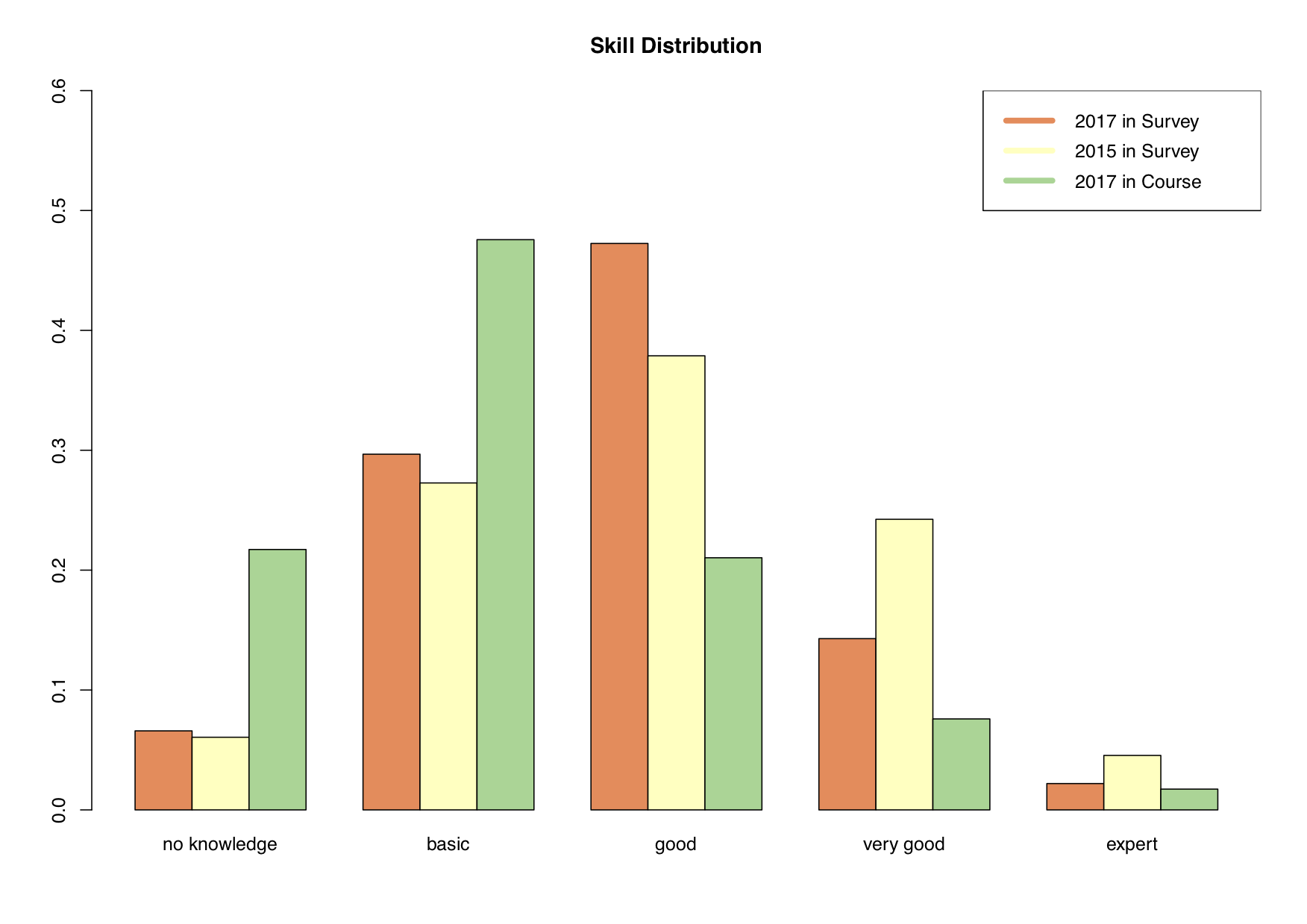}
    \caption{Skill Distributions in the 2017 Course and the Survey Groups}
    \label{pic:SkillDistribution}
\end{center}
\end{figure}

The first point that we examined further was the distribution of skills in Figure~\ref{pic:SkillDistribution} and Table~\ref{table:skill_levels}.
The skill level distribution of our survey responders from 2015 and 2017 is similar in general.
The only difference is that the in the group of responders from 2015 were some more participants with ``very good'' knowledge, whereas the responders from 2017 had some more responders with ``good'' knowledge.
Border options were mostly avoided within the survey.
When comparing those numbers with the  overall skill level distribution from the course audience of 2017, it becomes apparent that mostly participants with a higher skill level answered our survey.
The courses we conducted approached mainly beginners, therefore explains the high  but reasonable ratio of participants with ``no knowledge'' and the generally skewed distribution to the left of the overall course participants.

The outcomes of the graded questions meet our expectations with regard to most findings.
In general, the easier questions Q6 to Q8 show high mean scores for both surveyed groups.
The more difficult questions showed lower mean scores overall.

A second noticeable finding is that for most questions (8 of 10) the participants from 2017 scored better than those from 2015 in the survey questions.
Given the fact that the mean skill level of the 2015 survey participants was higher, we conclude that knowledge attrition indeed depends on the timespan between training and in this case surveying.
When comparing the results with the outcomes gathered during the course runtime of 2017, here labelled with ``former scores'', one might be surprised by the relatively large deltas for questions Q9 and Q10.
However, as stated before, questions Q9 and Q10 were asked before the actual course lectures were presented.
From that outcome we conclude that our participants acquired knowledge that they did not possess beforehand and are still able to recall it one or three years later.

For questions Q11 and Q12 we notice something different, the mean ``former scores'' are higher than the survey scores.
Q11 and Q12 were asked in a self-test directly after the respective lecture unit in the course 2017.
Therefore, we got mean scores of nearly 100\%.
It is only reasonable, that the mean scores dropped after some time.

The deltas grouped by skill levels and questions are relatively small, but add up in total.
We refrain from drawing conclusions for skill level 0 (``no knowledge'') and skill level 5 (``expert'') due to the low participant number.
Nonetheless, for skill levels basic to very good, it is noticeable that participants from 2017 achieved higher mean scores than those from 2015.
The shares come most prominently from the harder questions Q12 to Q15, suggesting that they are a suitable means to capture and distinguish knowledge gaps.

The mean score gap between the groups not having programmed after the course and those who had ($6.44$ vs. $8.47$ for 2017, $3.47$ vs. $7.32$ for 2015) further reassures us that application and recapitulation of knowledge leads to better retention.
The numbers also support the hypotheses that knowledge attrition happens over time, as the values for the participants of our course in 2015 are lower than those for the participants of 2017.

With regard to our research questions, we conclude that programming knowledge lessens noticeably over time [RQ1].
The skill level had only a weak correlation towards the survey scores.
With regards to knowledge attrition, we noticed no consistent effect depending on the skill level [RQ2].
Students that programmed in the meantime achieved a higher score in the graded survey questions in general, and our results show an even stronger effect for participants that took the course 3 years ago [RQ3].
Despite being a self-stated higher skilled group, our students from 2015 scored worse than those that took the course in 2017, leading to the conclusion that the passed timespan does have a measurable effect on knowledge attrition [RQ4].

\section{Limitations and Future Work}\label{sec:future-work}
General learnings from this experiment are mostly limited by the relatively low feedback rate we encountered. 
We could not go much deeper into uncovering specific misconceptions within specific areas (such as boolean logic or array access), as the total amount of answers stating a certain wrong option lead to a too small subgroups for further analysis of interdependence.
To improve this, we plan to survey the participants of our next MOOC after a period of six months already.
We will further stress out in that MOOC that constant repetition is very effective in fortifying knowledge and offer them an opt-in possibility to be mailed with additional training exercises and quizzes and self-tests after the course runtime.
 

 \section{Conclusion}\label{ch:conclusion}
There are many articles and blog posts, claiming ``MOOCs are here to stay''.
The important question, whether the knowledge acquired through MOOCs is here to stay, remains mostly unanswered as of now.
Existing literature mostly focusses on dropout rates, which yields little benefit in actually measuring the long-term impact of MOOCs.
On the basis of two programming MOOCs conducted in the years 2015 and 2017, we developed a survey consisting of 15 questions in order to gain some insight into the long-term retention of the conveyed knowledge.
Response rates were low, even for typical MOOC settings, as only a fraction of participants still had enough commitment to fill out the survey months after course completion without further external motivation. 
We shared the anonymized outcomes of the gathered data and were able to answer our four research questions regarding knowledge retention and attrition.
 Our findings serve as a starting point into exploring knowledge retention in programming MOOCs and can be further generalized in future work.
For our MOOCs, knowledge fades noticeably over time, however not at an alarming rate.
Especially basics seemed to be learned by heart.
Knowledge attrition did not depend on the skill level of the participant, and might be partly counteracted by applying the knowledge.
Determining optimal timespans for recapitulation remains an open research question, to be tackled in future course iterations.

\balance
\bibliography{retention-rates}

\begin{thebibliography}{10}

\bibitem{bawa_retention_2016}
Papia Bawa.
\newblock Retention in {Online} {Courses}: {Exploring} {Issues} and
  {Solutions}—{A} {Literature} {Review}.
\newblock {\em SAGE Open}, 6(1):2158244015621777, January 2016.

\bibitem{behling_translating_2000}
Orlando Behling and Kenneth~S. Law.
\newblock {\em Translating {Questionnaires} and {Other} {Research}
  {Instruments}: {Problems} and {Solutions}}.
\newblock SAGE, May 2000.

\bibitem{belleflamme_economic_2016}
Paul Belleflamme and Julien Jacqmin.
\newblock An {Economic} {Appraisal} of {MOOC} {Platforms}: {Business} {Models}
  and {Impacts} on {Higher} {Education}.
\newblock {\em CESifo Economic Studies}, 62(1):148--169, March 2016.

\bibitem{breslow_studying_2013}
Lori Breslow, David~E. Pritchard, Jennifer DeBoer, Glenda~S. Stump, Andrew~D.
  Ho, and Daniel~T. Seaton.
\newblock Studying {Learning} in the {Worldwide} {Classroom} {Research} into
  {edX}'s {First} {MOOC}.
\newblock {\em Research \& Practice in Assessment}, 8:13--25, 2013.

\bibitem{dale_audiovisual_1969}
Edgar Dale.
\newblock {\em Audiovisual methods in teaching.}
\newblock Dryden Press, New York, 1969.

\bibitem{fischer_revenue}
Helge Fischer, Stefan Dreisiebner, Oliver Franken, and Martin Ebner.
\newblock {Revenue} {vs}. {Costs} {of} {MOOC} {Platforms}. {Discussion} {of}
  {Business} {Models} {for} {XMOOC} {Providers}, {based} {on} {Empirical}
  {Findings} {and} {Experiences} {During} {Implementation} {of} {the} {Project}
  {IMOOX}.
\newblock page~11.

\bibitem{garrison_continuing_2007}
Julie Garrison.
\newblock Continuing {Education} and {Knowledge} {Retention}: {A} {Comparison}
  of {Online} and {Face}-to-{Face} {Deliveries}.
\newblock {\em Articles}, April 2007.

\bibitem{george_lean_2004}
Michael~L. George, John Maxey, David Rowlands, and Mark Price.
\newblock {\em The {Lean} {Six} {Sigma} {Pocket} {Toolbook}: {A} {Quick}
  {Reference} {Guide} to 100 {Tools} for {Improving} {Quality} and {Speed}}.
\newblock McGraw-Hill Education, New York, 1st edition, August 2004.

\bibitem{jones_optimal_2013}
W.~Paul Jones and Scott~A. Loe.
\newblock Optimal {Number} of {Questionnaire} {Response} {Categories}.
\newblock {\em SAGE Open}, 3(2):2158244013489691, April 2013.

\bibitem{kirkpatrick_evaluating_1994}
Donald~L. Kirkpatrick.
\newblock {\em Evaluating {Training} {Programs}: {The} {Four} {Levels}}.
\newblock Berrett-Koehler, San Francisco: Emeryville, CA, November 1994.

\bibitem{KizilcecAttrition2015}
Ren{\'e}~F. Kizilcec and Sherif Halawa.
\newblock Attrition and achievement gaps in online learning.
\newblock In {\em Proc. L@S '15}, pages 57--66, New York, NY, USA, 2015. ACM.

\bibitem{kruger_unskilled_1999}
Justin Kruger and David Dunning.
\newblock Unskilled and unaware of it: {How} difficulties in recognizing one's
  own incompetence lead to inflated self-assessments.
\newblock {\em Journal of Personality and Social Psychology}, 77(6):1121--1134,
  1999.

\bibitem{moocmaker2017}
MOOC-Maker.
\newblock Construction of management capacities of moocs in higher education.

\bibitem{morrison_2013_programming_age}
Patrick Morrison and Emerson Murphy-Hill.
\newblock Is programming knowledge related to age? an exploration of stack
  overflow.
\newblock In {\em Proceedings of the 10th Working Conference on Mining Software
  Repositories}, MSR '13, pages 69--72, Piscataway, NJ, USA, 2013. IEEE Press.

\bibitem{naidr_long_term_2004}
J.~P. Naidr, T.~Adla, A.~Janda, J.~Feberova, P.~Kasal, and M.~Hladikova.
\newblock Long-{Term} {Retention} of {Knowledge} {After} a {Distance} {Course}
  in {Medical} {Informatics} at {Charles} {University} {Prague}.
\newblock {\em Teaching and Learning in Medicine}, 16(3):255--259, July 2004.

\bibitem{phillipsRetention2015}
Alana~S. Phillips.
\newblock Retention: {Course} {Completion} {Rates} in {Online} {Distance}
  {Learning}, December 2015.

\bibitem{phillips_value_2007}
Jack~J. Phillips and Patricia~Pulliam Phillips.
\newblock {\em The {Value} of {Learning}: {How} {Organizations} {Capture}
  {Value} and {ROI} and {Translate} {It} into {Support}, {Improvement}, and
  {Funds}}.
\newblock John Wiley \& Sons, August 2007.

\bibitem{renz_opensap_2016}
Jan Renz, Florian Schwerer, and Christoph Meinel.
\newblock {openSAP}: {Evaluating} {xMOOC} {Usage} and {Challenges} for
  {Scalable} and {Open} {Enterprise} {Education}.
\newblock {\em International Journal of Advanced Corporate Learning (iJAC)},
  9(2):34--39, August 2016.

\bibitem{smith_six_sigma_1993}
Bill Smith.
\newblock Six-sigma design (quality control).
\newblock {\em IEEE Spectrum}, 30(9):43--47, September 1993.

\bibitem{Whitehill2015Beyond}
Jacob Whitehill, Joseph~Jay Williams, Glenn Lopez, Cody~Austun Coleman, and
  Justin Reich.
\newblock Beyond prediction: First steps toward automatic intervention in mooc
  student stopout.
\newblock {\em Available at SSRN 2611750}, 2015.

\bibitem{yudelson_investigating_2014}
Michael Yudelson, Roya Hosseini, Arto Vihavainen, and Peter Brusilovsky.
\newblock Investigating {Automated} {Student} {Modeling} in a {Java} {MOOC}.
\newblock In {\em Educational {Data} {Mining} 2014}, pages 261--264, London,
  UK, July 2014. University of Pittsburgh.

\bibitem{zabrucky_testing}
Karen~M. Zabrucky and Rebecca~B. Bays.
\newblock Improving students' retention of classroom material through the
  testing effect.
\newblock {\em College Teaching}, 63(2):91--91, 2015.

\end{thebibliography}
\bibliographystyle{plain}

\end{document}